\documentclass{ws-procs975x65}

\def\beq{\begin{equation}}
\def\eeq{\end{equation}}

\begin{document}

\title{Development of superconducting Klystron cavity for the Mario Schenberg gravitational wave detector}

\author{V. Liccardo}

\address{ITA-Instituto Tecnol\'ogico de Aeron\'autica, S\~ao Jos\'e dos Campos - SP - Brasil\\
E-mail: vic2000@hotmail.it\\}

\author{O. D. Aguiar$^*$ and E. K. de Fran\c{c}a}

\address{INPE-Instituto Nacional de Pesquisas Espaciais,  S\~ao Jos\'e dos Campos - SP - Brasil\\
$^*$E-mail: odylio.aguiar@inpe.br,
E-mail: ekfranca@gmail.com}

\begin{abstract}
Superconducting reentrant cavities can be used in parametric transducers for Gravitational Wave antennas. The Mario Schenberg detector, which is being built by the GRAVITON group at Instituto Nacional de Pesquisas Espaciais (INPE), basically consists of a resonant mass (ball) and a set of parametric transducers in order to monitor the fundamental modes of vibration. When coupled to the antenna, the transducer-sphere system will work as a mass-spring system. In this work the main task is the development of parametric transducers consisting of reentrant superconducting cavity with high performance to be implemented in the Mario Schenberg detector. Many geometries, materials and designs have been tested and compared to optimize parameters such as electric and mechanical Q-factor. The aim is the construction of a complete set of nine parametric transducers that, attached to the spherical antenna, will possibly reach the sensitivity $h$ $\sim$ 10$^{-22}$ $Hz$$^{-1/2}$ in the near future.
\end{abstract}

\keywords{Resonant detectors; Gravitational waves; Parametric transducer; Klystron resonant cavities; GW antennas.}

\bodymatter

\section{INTRODUCTION}

\begin{figure}[!h]
   \begin{center}
   \includegraphics[scale=0.25]{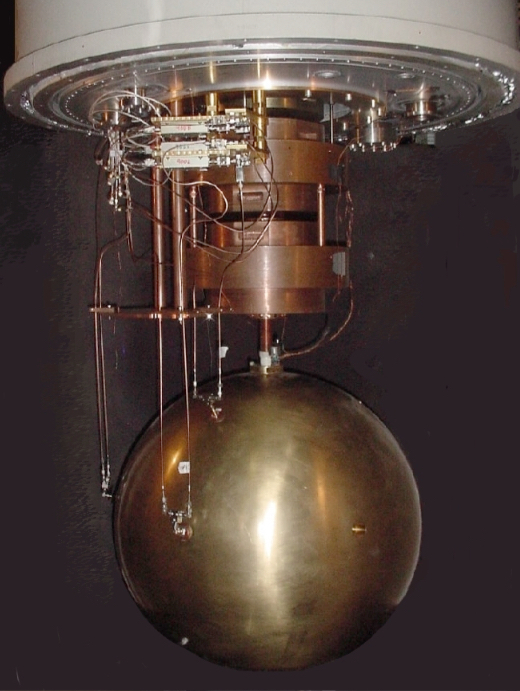}
    \includegraphics[scale=0.265]{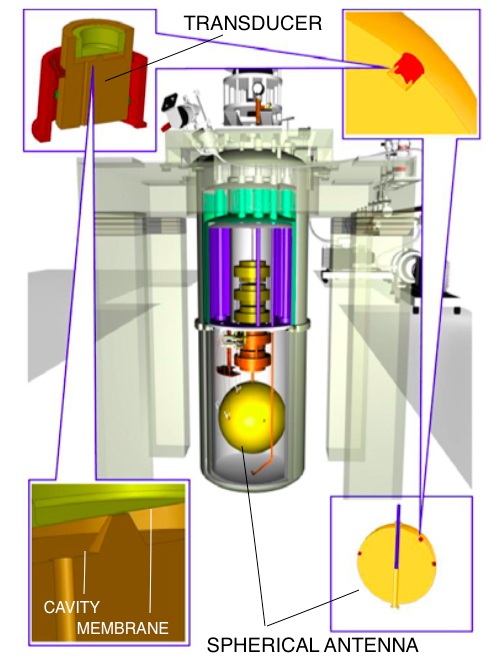}
     \caption{\footnotesize{Picture of the metal sphere.}}
     \label{sphere}
\end{center}
\end{figure}

The Mario Schenberg gravitational wave detector is based on the detection of five quadrupole modes relative to the mechanical vibrations of a 65 cm diameter spherical resonant mass of 1150 kg (Fig. \ref{sphere}). In the case of bars, like interferometers, only one mode is monitored. The solid metal sphere of Mario Schenberg antenna has a high mechanical quality factor $Q$ $\sim$ 10$^{7}$ which will allow an isotropic and a multichannel analysis of signal. Its resonant frequency is $f_{0}$ $\sim$ 3.2 $kHz$. The modes of oscillation can act as independent antennas oriented in different directions. To monitor the five quadrupolar modes a set of transducers are coupled with sphere allowing the GW detection \cite{aguiar}.

Transducers are employed to convert mechanical into electrical energy, i.e. mechanical vibrations are converted into an electrical signals. In Schenberg detector, it was decided to use parametric transducers with microwave klystron-type cavities \cite{blair}. The klystron cavities, which make up part of the transducer, are ``pumped" with AC signals (10 GHz), which is modulated by the resonant cavity. A feature of the parametric transducer is occurring in a power amplifier transduction process, called parametric gain \cite{tobar}.

The Mario Schenberg detector transduction system consists of nine transducers fixed on the surface of the sphere (Fig. \ref{sphere}). Their respective electronic amplification, demodulation and digitization will operate independently. To prevent seismic noise to reach the antenna through the wiring of the transducers, the electrical connection between the transducers and the cabling are made through wireless coupling. This coupling must have the lowest possible loss in the resonance frequency region of the cavity.

%

\section{REENTRANT CAVITY DESCRIPTION}

\begin{figure}[!h]
   \begin{center}
   \includegraphics[scale=0.33]{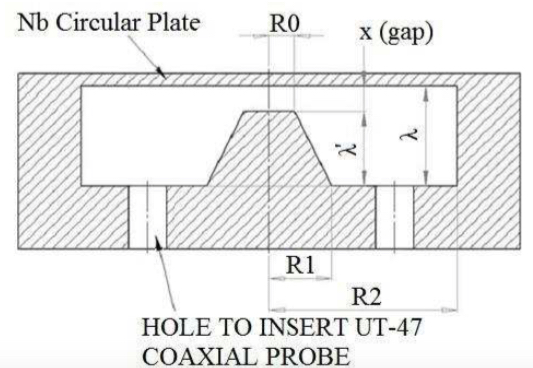}
     \caption{\footnotesize{Sketch of the klystron cavity used for the parametric transduction system of the Mario Schenberg detector.}}
     \label{cavity}
\end{center}
\end{figure}

The cylindrical klystron (reentrant) cavity is made of niobium, featuring a central conical post and closed on the top by a metallic circular membrane which forms a narrow axial gap with the central post (Fig. \ref{cavity}). 

The gap spacing is a key parameter since the cavity resonant frequency depends very strongly on it. The cavity may be considered like an LCR resonator whose capacitance is determined by the gap spacing \cite{pimentel}. The cavity capacitance is modulated by the motion of the plate (membrane) with respect to the gap providing a capacitive effect on the frequency tuning and allowing to employ it as part of electromechanical transducer. The transducer-sphere system works as a mass-spring system with three modes, where the first one is the antenna effective mass, the second one is constituted by the mechanical structure of the transducer, and the last one being the membrane itself which closes the transducer microwave cavity and modulate it around 3.2 $kHz$.

The main characteristics of a resonant cavity is the resonance frequency ($f_{0}$) and the quality factor ($Q$). The resonant frequency is given by:

\begin{equation}
{f_0} = \frac{{{k_0}}}{{2\pi }}\
\end{equation}

where $k_{0}$=2$\pi$/$\lambda_{0}$ is the wave number in the resonance condition, which depends on the geometry of the cavity and on the excited mode. The $Q$ is an important parameter to characterize the resonant cavities. The factor determines the performance of the resonant circuit being proportional to the ratio between the energy stored and lost in the circuit per cycle. It is also related to the width of the passband of a generic system by:

\begin{equation}
Q = 2\pi \frac{{Energy{\text{ }}Stored}}{{Energy{\text{ }}Dissipated{\text{ }}per{\text{ }}Cycle}} = \frac{{{\omega _0}}}{{\Delta \omega }}\
\end{equation}

where $\omega_{0}$ is the system resonance frequency and $\Delta$$\omega$ is the frequency bandwidth for a half power attenuation.

The detector sensitivity strongly depends on several types of noise, being the transducer series noise one of the most crucial \cite{costa}. Transducer noise is affected by the cavity characteristic parameters such as the displacement sensitivity ($df$/$dx$) (variation of the frequency as function of the gap spacing $x$), the unloaded electrical Q-factor ($Q_{0}$) and the electromagnetic coupling ($\beta$) \cite{paula2}.

Turns out to be clear that the main goal in the cavity development is to maximize the ($df$/$dx$) and the $Q_{0}$, so that the noise produced by the transducer system is sufficiently low, and mechanical amplification of the gravitational wave signal via a multi-mode transducer is sufficient.

\section{EXPERIMENTAL AND DEVELOPMENT METHOD}\label{devel}

\begin{figure}[!h]
   \begin{center}
   \includegraphics[scale=0.35]{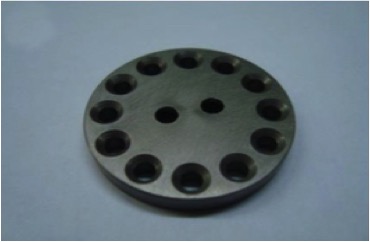}
    \includegraphics[scale=0.35]{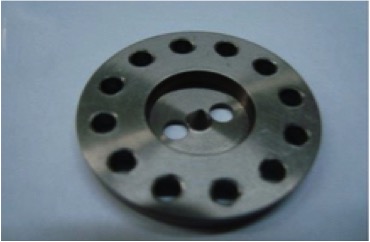}
     \caption{\footnotesize{Niobium reentrant cavity sample.}}
     \label{kly}
\end{center}
\end{figure}

The purpose of the following work is to manufacture niobium reentrant cavities (Fig. \ref{kly}) to be implemented on transducers, and study the effect of the niobium nitride (NbN) on the $Q_{0}$. 10 niobium transducers were machined at INPE featuring 28 mm diameter and 4 mm thickness. The transducer head consists of the niobium membrane along with the cover, both bonded to the klystron cavity. 12 holes were drilled at the edges for the system closing.

The resonance frequency of the membrane should be 3.2 kHz, the same frequency of the sphere quadrupole modes. 
The cavities were made completely of niobium and are electromagnetically resonant at $\sim$ 9,44 GHz, which is the same frequency produced by the external injection signal source, an oscillator with ultra-low phase noise. 

The key to obtain transducers with the best performance is the adjustment of the microwave cavity in such a way to resonate at its frequency. In theory, this will be achieved when the gap spacing is close to 3 $\mu$m. In order to develop cavities with high Q-factor the samples went through several processes which are summarized in the steps described below.

\subsection{Sanding and Polishing}

The method used to decrease the gap spacing was to manually sand the circular cover surface which is in contact with the body of the transducer, except for the central conical post. Thanks to the sanding the thickness of the sample decreases making the gap between the post and the surface smaller. As result the post approaches the membrane, since the latter is placed on the cover surface.

\begin{figure}[!h]
   \begin{center}
   \includegraphics[scale=0.05]{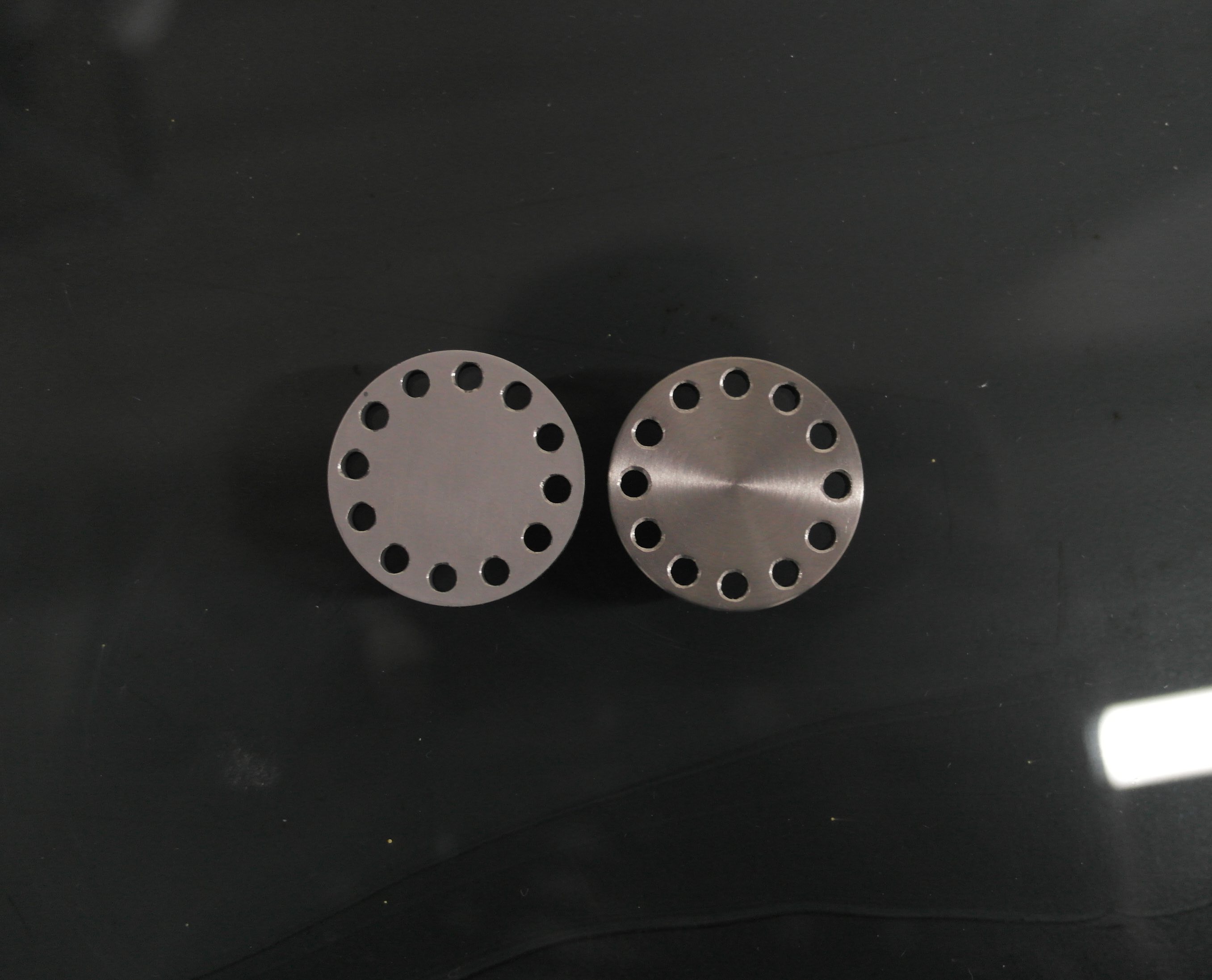}
     \caption{\footnotesize{Samples of niobium cavities before ($left$) and after ($right$) the sanding/polish process.}}
     \label{polish}
\end{center}
\end{figure}

The abrasives used for the sanding process were first cut, in the form of tapes, from commercial sandpapers with granulations of 600, 1200 and 2000. The method employed consists in sanding the outer part of the niobium cylindrical cavity (the top of the post remained untouched) and the  circular niobium cover until the surface got shiny. The technique adopted to obtain a shining surface relies in sanding the cavity with diagonal and circular movements (``eight'' motion), using sandpapers in descending order of granulations. We obtained samples with very good polished surfaces (Fig. \ref{polish}) with gaps being reduced to the expected value.
Once polished the cavities suffered a chemical cleaning in the ultrasound to remove the oxide and the material grain from the surfaces, being now ready for the subsequent step.

\subsection{Implantation}

At room temperature, the natural oxidation of niobium ($Nb_{2}$$O_{5}$) in the presence of air is slow but can reach in only a few days 6 nm of thickness \cite{grundner}. The formation of the natural oxide greatly reduces the $Q_{0}$ of the cavity. In order to avoid the natural oxidation, it is necessary to apply on the cleaned sample surface a process such that the interested area becomes passive. Although anodization \cite{martens} is the most common technique, implantation of NbN was chosen to prevent the oxidation of the niobium cavities \cite{oliveira}. NbN behaves as a superconductor at low temperatures and shows excellent resistance to oxidation. It oxidizes in air only from 800 $^{\circ}$C.

\begin{figure}[!h]
   \begin{center}
   \includegraphics[scale=0.28]{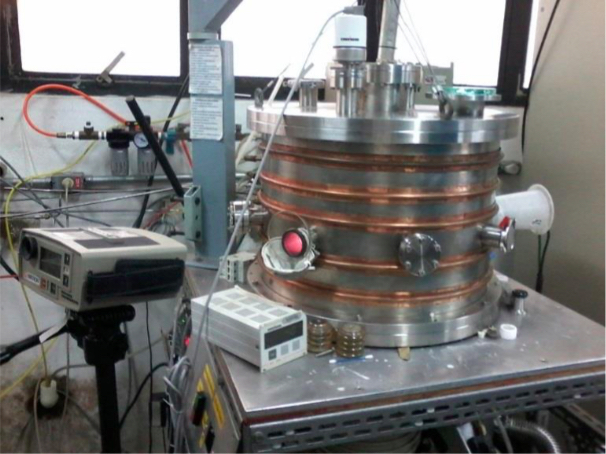}
     \caption{\footnotesize{High temperature chamber for the 3IP process.}}
     \label{3IP}
\end{center}
\end{figure}

The process adopted was the ion implantation by immersion in plasma (3IP). The sample to be treated is initially immersed in a nitrogen plasma and is polarized with a negative high voltage pulse. The applied high negative voltage accelerates the plasma electrons away from the sample whereas ions are accelerated toward it. The result  is the creation of a plasma layer around the sample surface which allows the implantation of nitrogen ions. NbN layers of $\sim$ 4 $\mu$m can be obtained on high purity niobium samples with the described process \cite{oliveira}.

\subsection{Frequencies Tuning}

Once the samples have been processed, the cavities were mounted to check if the electric resonance frequencies are within the expected range.
However, adjust the cavities to resonate at a specific frequency is a major challenge of the commissioning phase since the gap spacing needs to be set to the order of microns. Moreover, has been shown that the resonance frequency does not change linearly with the gap  \cite{ferreira}.

\begin{figure}[!h]
   \begin{center}
   \includegraphics[scale=0.05]{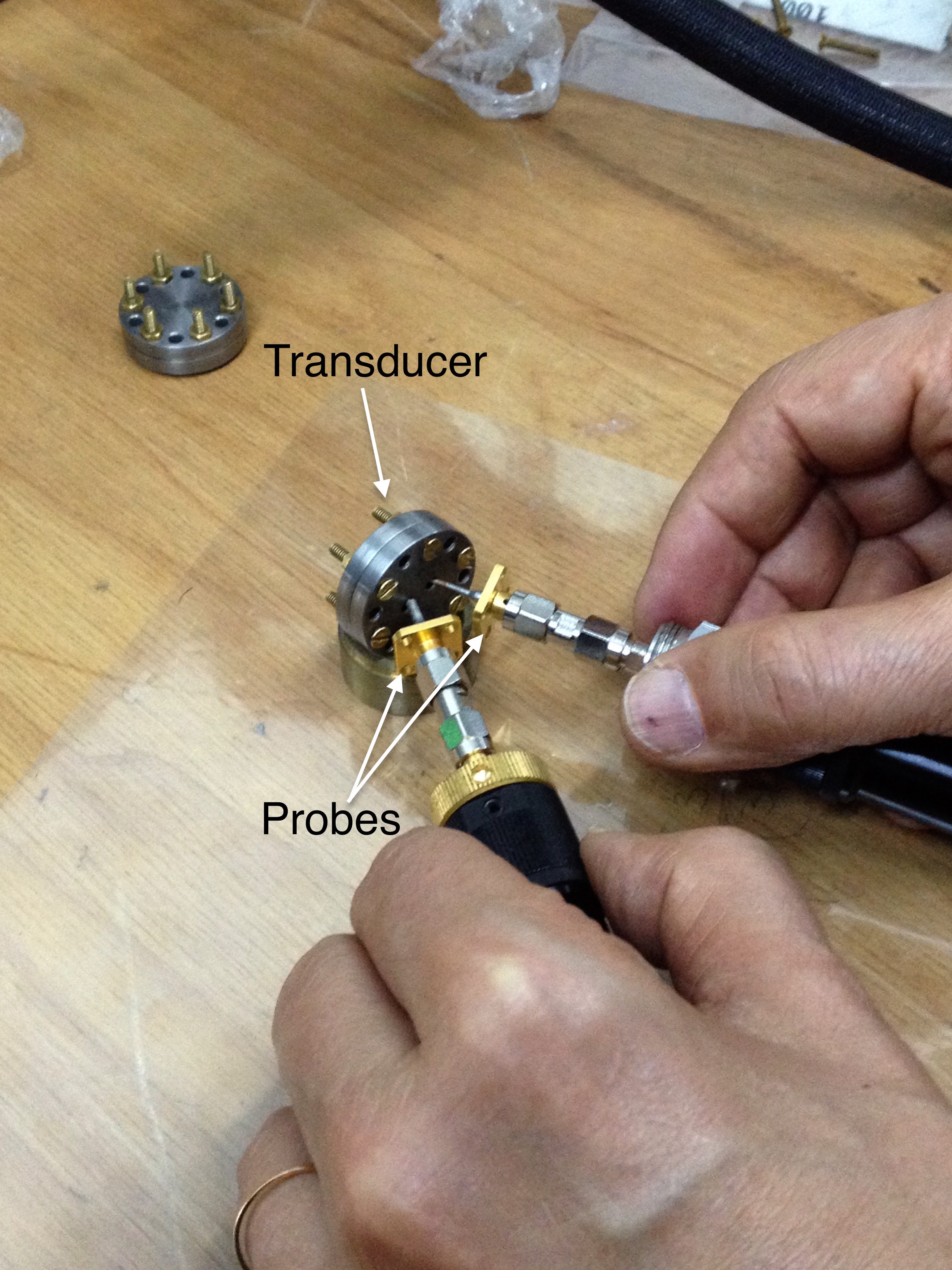}
   \includegraphics[scale=0.05]{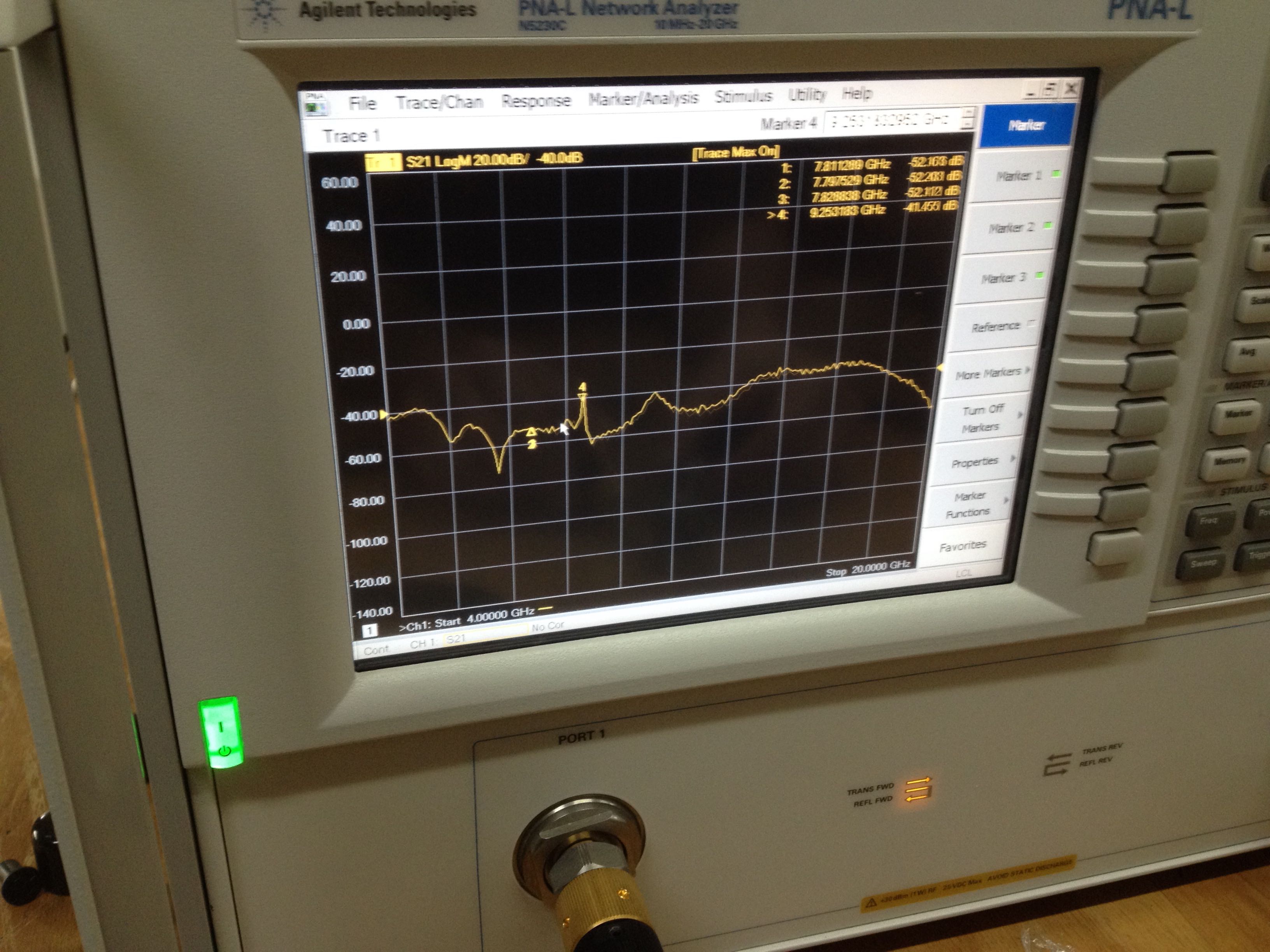}
     \caption{\footnotesize{$Left$: Resonance frequency measurements of a cavity after the lapping/implantation process. $Right$: The vector network analyzer.}}
     \label{freq}
\end{center}
\end{figure}

The resonant frequency measurements were accomplished with the vector network analyzer, in transmission mode, by inserting two probes into the cavity (Fig. \ref{freq}). Through a fine lapping, using sandpaper with small granulation, was possible to adjust the central post size in such a way to regulate the cavity gap spacing and tune the desired resonant frequency. With the same method also the niobium cover thickness was monitored with a micrometer after each thinning process. Although the gap is very small ($\sim$ 3 $\mu$m) and the frequency depends very strongly on the gap, the results obtained show that the assembling of the membrane is fairly reliable and the gap spacing is reasonably reproducible.

The next step consists in the cryogenic Q-factor measurements, since niobium becomes a superconductor for any temperature below the critical temperature (9.2 $K$). Performing measurements at cryogenic conditions will thus allow to obtain high quality factors.

\section{Q-FACTOR MEASUREMENTS}

\begin{figure}[!h]
   \begin{center}
   \includegraphics[scale=0.3]{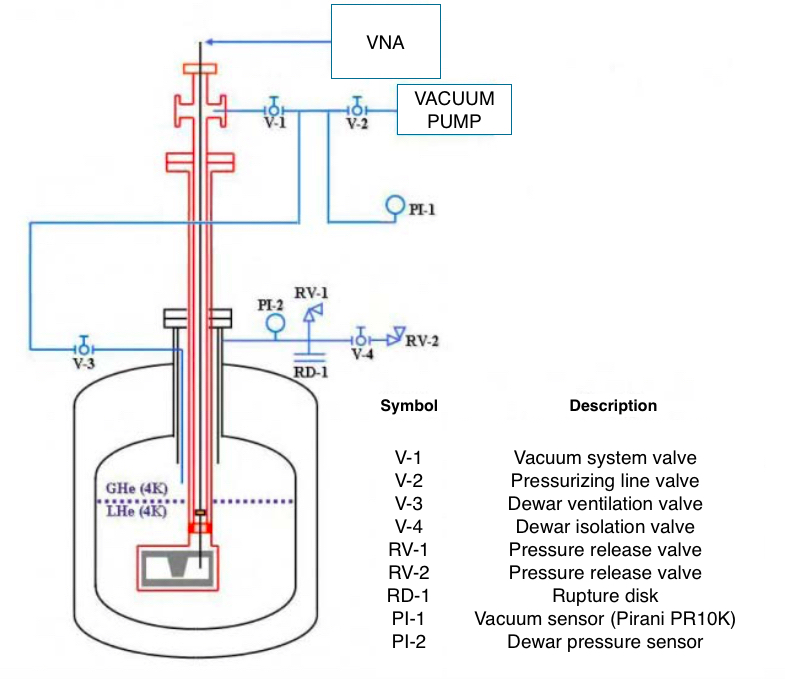}
     \caption{\footnotesize{Diagram of the experimental setup used to carry out cryogenic measurements of Q-factor of reentrant cavities.}}
     \label{appar}
\end{center}
\end{figure}

The cryogenic Q-factor measurements, will be carried out using the experimental apparatus existing in the GRAVITON laboratory (Fig. \ref{appar}). More details on the facility are reported elsewhere \cite{paula, paula2}.

For the experiments under superconductivity conditions of Niobium ($T$$<$9.2 $K$), will be used a cryostat already present in the laboratory, enabling the cavities cooling through the insertion of them inside a dewar containing liquid helium (LHe). The measurements of resonance frequency, electromagnetic coupling and  $Q_{0}$  in reflection mode will be performed according to this assembly using a vector network analyzer (Agilent PNA-L) which has a scanning font operating in the frequency band of 0.01-20 GHz and measurement capability of insertion loss, gain and return loss. A scanning signal, in a predefined frequency range, is injected into the cavity and analyzed in reflection mode, being absorbed in the resonant region of the klystron mode.

\section{CONCLUSIONS}

We have already succeeded to govern the implantation 3IP process, obtaining high quality NbN layer for protecting the inner surfaces of the cavities against progressive oxidation. The sanding method for the adjustment (mostly decrease) of the cavities gap spacing showed to be effective and reproducible since the cavities frequencies showed values close to those expected. The upcoming Q-factor run tests will allow us to understand the effect of the NbN implantation and to support the improvement of the transducers performance, which play a key role in the gravitational wave antenna sensitivity.

A detailed description of Q-factor measurements, results and further developments can be found in an incoming dedicated paper.

\section*{Acknowledgments}

I, Vincenzo Liccardo acknowledge Prof. Manuel Malheiro for kindly give support to this work through the ``Science Without Borders" project from CAPES. We would like to thank INPE/MCTI and the Department of Materials Physics and Mechanics, Institute of Physics, University of Sao Paulo.

\end{document}